\documentclass[journal]{IEEEtran}

\usepackage[T1]{fontenc}
\usepackage{tabularx}
\usepackage{pdflscape}
\usepackage{multirow}
\usepackage{multicol}
\usepackage{upquote}
\usepackage[bookmarks=false]{hyperref}
\usepackage{cite}
\usepackage[utf8x]{inputenc} 
\usepackage{longtable}
\usepackage{colortbl}
\usepackage{tikz}
\usepackage{caption}
\usepackage{xcolor}
\usetikzlibrary{arrows,shapes,positioning,shadows,trees}
\usepackage{rotating, adjustbox}
\usepackage{pgfplots}
\usepgfplotslibrary{groupplots}
\pgfplotsset{compat=1.12}
\usepackage{bbold}
\usepackage{tkz-euclide,subfigure}
\usetikzlibrary{patterns}
\usetikzlibrary{knots,calc}
\usepackage{algorithm}
\usepackage{algpseudocode}
\usepackage{supertabular} 
\usepackage{graphicx}
\usepackage{lineno,hyperref}
\usepackage{amsmath,amsfonts,amssymb}
\usepackage{booktabs}
\usepackage{flushend}
\usepackage{parskip}
\usepackage{smartdiagram}
\def\BibTeX{{\rm B\kern-.05em{\sc i\kern-.025em b}\kern-.08em T\kern-.1667em\lower.7ex\hbox{E}\kern-.125emX}}
\usepackage{pifont}
\usepackage{multirow}
\usepackage{booktabs}
\usepackage{tabularx}
\usepackage{float}
\usepackage[vlines]{tabularht}
\usepackage{rotating}
\usepackage{array,multirow,makecell}
\modulolinenumbers[5]
\usepackage{lscape}
\usepackage[cmintegrals]{newtxmath}
\usepackage{tabularx}
\usetikzlibrary{shadows}
\usepackage{forest}
\usepackage{fontawesome}
\usepackage{comment}
\usepackage{diagbox}
\usepackage{caption}

\definecolor{color0}{rgb}{1.000, 1.000, 1.000} 
\definecolor{color1}{rgb}{1.000, 1.000, 0.800}  
\definecolor{color2}{rgb}{0.698, 0.875, 0.541}
\definecolor{color3}{rgb}{0.400, 0.741, 0.741}
\definecolor{color4}{rgb}{0.094, 0.506, 0.749}

\definecolor{morange}{rgb}{0.8,0.2,0}
 
\IEEEoverridecommandlockouts

\definecolor{Gray}{gray}{0.9}
\newcolumntype{g}{>{\columncolor{Gray}}p}

\usepackage{etoolbox}

\newcommand{\PreserveBackslash}[1]{\let\temp=\\#1\let\\=\temp}
\newcolumntype{C}[1]{>{\PreserveBackslash\centering}p{#1}}

\let\mybibitem\bibitem
\renewcommand{\bibitem}[1]{%
\ifstrequal{#1}{rathee2021design}{\color{blue}\mybibitem{#1}}
{\ifstrequal{#1}{barka2021sthm}{\color{blue}\mybibitem{#1}}
{\color{black}\mybibitem{#1}}}%
}

\usepackage[inline]{enumitem}

\begin{document}

\title{MalVol-25: A Diverse, Labelled and Detailed Volatile Memory Dataset for Malware Detection and Response Testing and Validation}

\author{Dipo~Dunsin*, Mohamed~Chahine~Ghanem, Eduardo Almeida Palmieri
\thanks{------------------------------------------------------------------------------------------}
\thanks{*~Dr Dipo Dunsin is the corresponding author. email: \href{d.dunsin@londonmet.ac.uk}{d.dunsin@londonmet.ac.uk}| Dr. D.~Dunsin and  Mr. E.~Almeida Palmieri are with the Cyber Security Research Centre, London Metropolitan University, London, UK. Dr. M.C.~Ghanem is with Cybersecurity Institute, Department of Computer Science, The University of Liverpool, Liverpool, UK.}

}

\maketitle

\begin{abstract}
This paper addresses the critical need for high-quality malware datasets that support advanced analysis techniques, particularly machine learning and agentic AI frameworks. Existing datasets often lack diversity, comprehensive labelling, and the complexity necessary for effective machine learning and agent-based AI training. To fill this gap, we developed a systematic approach for generating a dataset that combines automated malware execution in controlled virtual environments with dynamic monitoring tools. The resulting dataset comprises clean and infected memory snapshots across multiple malware families and operating systems, capturing detailed behavioural and environmental features. Key design decisions include applying ethical and legal compliance, thorough validation using both automated and manual methods, and comprehensive documentation to ensure replicability and integrity. The dataset’s distinctive features enable modelling system states and transitions, facilitating RL-based malware detection and response strategies. This resource is significant for advancing adaptive cybersecurity defences and digital forensic research. Its scope supports diverse malware scenarios and offers potential for broader applications in incident response and automated threat mitigation.
\end{abstract}

\begin{IEEEkeywords}
Malware, Ransomware, RAM, Volatile Memory, Incident Response, Artificial Intelligence (AI), Digital Forensics, Cyber Attacks.
\end{IEEEkeywords}

\section{Introduction}
\subsection{Overview of Machine Learning and AI in Cybersecurity}
Malware threats increasingly challenge cybersecurity, demanding timely detection and mitigation to protect critical systems \cite{Malik}. Traditional methods struggle to keep pace with rapidly evolving variants, resulting in persistent vulnerabilities and risks \cite{Or-Meir}. Consequently, there is growing interest in advanced computational techniques like machine learning and agentic AI to improve malware analysis and incident response \cite{Dunsin}. The effectiveness of such approaches depends fundamentally on high-quality, representative datasets for training and evaluation. Existing datasets often lack diversity, contain outdated samples, and suffer from insufficient labelling \cite{Anderson}. Public malware datasets frequently fall short of the volume and complexity needed for machine learning, which relies on dynamic interactions with diverse environments to optimise policies \cite{Dunsin}. This creates an urgent need for systematically generated datasets that capture multifaceted malware behaviours in realistic contexts. Current datasets mostly focus on static malware features or conventional ML models, leaving machine learning and agentic AI applications underexplored. The scarcity of well-organised datasets tailored to machine learning and agentic AI hampers advances in automated detection and response systems \cite{Dunsin}. This research addresses this gap by proposing a systematic malware dataset generation method designed specifically for machine learning and agentic AI frameworks. It prioritises diversity, adaptability, and realism to accurately reflect real-world malware scenarios. The dataset approach integrates automated malware execution with dynamic monitoring, producing rich datasets featuring behavioural and environmental aspects critical to incident response \cite{Dunsin}. The method supports the development of adaptive, resilient detection systems capable of real-time response to emerging threats.

ML and AI have revolutionised cybersecurity by enabling automated detection and responses based on learnt data patterns. ML encompasses supervised, unsupervised, and reinforcement learning, each suited to distinct challenges. Unlike traditional signature-based methods that often fail against novel or polymorphic malware, ML models adapt continuously, identifying subtle anomalies and new threats. Consequently, AI-powered systems have shifted defences towards intelligent, adaptive mechanisms that operate beyond static rules. The paper proceeds as follows: \textbf{\textit{Section}}\ref{sec:B} reviews related malware datasets and RL applications, \textbf{\textit{Section}}\ref{sec:C} describes the dataset generation methodology, \textbf{\textit{Section}}\ref{sec:D} presents evaluation results, and \textbf{\textit{Section}}\ref{sec:I} concludes with future research directions.

\subsection{Research Aim and Objectives}
This research aims to address gaps in current malware datasets that hinder machine learning and agentic AI frameworks in malware analysis. It suggests a structured way to create various realistic datasets by combining automated malware testing setups with monitoring tools that work on different operating systems. The objectives include developing a robust approach to creating datasets that reflect real-world infection scenarios with detailed behavioural features, validating their quality through forensic analyses, documenting the data collection process for transparency, and demonstrating the dataset’s practical use in advancing adaptive malware detection and real-time incident response.

\subsection{Research Contributions}
This study contributes a novel dataset generation approach tailored to machine learning and agentic AI applications in malware detection. It provides a secure experimental environment with diverse malware samples and operating system variants, producing well-validated memory snapshots both before and after infection. The research offers comprehensive documentation and data integrity measures that ensure reproducibility and trustworthiness. Additionally, it highlights the dataset’s potential to enable advanced machine learning and agentic AI-based cybersecurity solutions by modelling system states and decision-making processes. These contributions collectively advance malware analysis, support forensic investigations, and provide a valuable resource for ongoing research and innovation in adaptive malware detection.

\subsection{Comparison with Existing Literature Reviews}
Compared to existing work, this research methodology demonstrates significant improvements in several key areas. While previous studies such as FabIoT \cite{Huertas1} and CMD\_2024 \cite{Nguyen} focused on specific environments like IoT or cloud systems with limited dynamic data and feature diversity, our approach employs a comprehensive virtual machine infrastructure that enables detailed monitoring across multiple operating systems. Our approach results in a richer and more diverse dataset, which captures a wide range of malware behaviours more effectively. Unlike MALVADA \cite{Raducu} and datasets like MalVis, which struggle with complicated data or missing information, our research offers clearly labelled memory snapshots of both clean and infected states, improving forensic, machine learning, and agentic AI applications. Careful malware selection and infection management ensure reliable, high-quality data, addressing consistency issues found in other studies. Additionally, our methodology stresses ethical and legal compliance, which was often overlooked in previous research. Combining automated tools with manual validation, we deliver a reliable, accurate dataset supporting advanced cybersecurity research. Finally, this adaptable and ethically responsible dataset approach surpasses many current datasets in diversity, detail, and practical value, making it a valuable resource for improving malware detection and response strategies.

\section{Related work}
\label{sec:B}
\subsection{Dataset Creation for Behaviour Modelling in IoT Malware}
Existing research on IoT malware detection relies mainly on datasets focusing on network traffic analysis. Common datasets like LITNET-2020 \cite{Damasevicius1}, IoT-23 \cite{Abdalgawad1}, and N-BaIoT \cite{Abbasi1} capture packet flows and network signatures, but they often ignore internal system behaviours such as CPU usage, memory access, and hardware performance counters. These datasets, typically generated by network monitoring tools or honeypots, limit representations of comprehensive malware behaviours on resource-constrained devices. Many contain outdated malware samples and seldom include zero-day or advanced persistent threats, which reduces effectiveness against evolving threats. Models trained on them often fail to generalise and miss nuanced malware actions inside IoT devices. To address this, Huertas et al. \cite{Huertas1} created FabIoT, collecting 72 system-level metrics from a Raspberry Pi under healthy and infected states. This enables a deeper understanding of malware behaviour at the device level, aided by anomaly detection tailored to IoT constraints. However, FabIoT’s reliance on one device and three malware types limits generalisability, and preprocessing raw data is challenging. Still, it marks an important advance.

\subsection{Advances and Challenges in Cloud Malware Datasets}
Malware detection research relies on diverse datasets that differ in scope, features, and collection methods. Traditional datasets often focus on static features like file metadata or dynamic features such as system call sequences, but this separation limits capturing malware behaviour, especially in complex cloud environments. To address this, Nguyen et al., \cite{Nguyen} introduced the CMD\_2024 dataset, a hybrid resource combining 12 static and 227 dynamic features from virtual machine introspection. This approach extracts cloud-specific behaviours that are otherwise difficult to monitor, offering a holistic view of malware's characteristics. CMD\_2024 contains over 20,000 samples covering various malware types, supporting binary and multi-class classification. Its open availability promotes reproducibility, addressing limitations of prior proprietary or small-scale datasets. However, challenges remain, including high feature dimensionality causing computational costs and class imbalances affecting detection of rare malware. Although CMD\_2024 enhances cloud-based malware datasets, continuous efforts are required to refine features and refresh samples in response to changing threats.

\subsection{Malware Detection via Sandbox Data}
Singh, Ikuesan, and Venter \cite{Singh} provide rich behavioural data by executing malware in controlled environments, extracting features like API calls, PE section entropy, and memory usage. Their study analysed over 3000 ransomware and benign samples via sandbox reports, producing datasets that supported high-accuracy detection models. However, dataset creation is laborious and non-reproducible, requiring manual execution and feature curation. Public platforms like Kaggle offer limited access, usually only processed datasets, restricting transparency and customisation. To address this, Singh, Ikuesan, and Venter \cite{Singh} introduced the MalFe platform, a community-driven repository for sharing and parsing raw sandbox reports, streamlining dataset generation, and promoting repeatability. Despite its advances, MalFe relies on user-submitted parsing scripts of varying quality, and wider adoption is needed for dataset diversity. Future enhancements, such as automated model training, could increase its impact. This research highlights that accessible raw data and collaborative tools are vital for overcoming malware dataset creation challenges and advancing detection research.

\subsection{MALVADA: Next-Gen Malware Datasets}
Malware research depends on high-quality datasets that accurately represent malicious behaviours. Existing datasets often contain simplified APIs or system call sequences that lack essential context, like parameters and return values, which limits deep behavioural analysis and AI detection. AWSCTD \cite{Ceponis} provides anonymised system call sequences but misses important details and uses broad malware family labels, reducing their usefulness. In contrast, the MALVADA framework generates execution trace datasets enriched with detailed context such as process trees, API parameters, resource accesses, and synchronisation objects, as described by Raducu et al., \cite{Raducu}. Its modular architecture enables the creation of large datasets with minimal user effort, exemplified by WinMET’s roughly 10,000 richly annotated malware traces. WinMET improves label accuracy by integrating advanced classification tools like AVClass. However, challenges include malware diversity, resource-intensive trace generation, and limited accessibility. While the MALVADA advances dataset quality, continuous updates and collaborative efforts remain essential to address evolving malware variants and expand coverage.

\subsection{Advances in Android Malware Image Datasets}
The MalVis dataset, \cite{Makkawy}, offers over 1.3 million RGB images from bytecode visualisations combining entropy and N-gram analyses. This captures structural anomalies and obfuscation patterns like encryption, packing, and compression, which simpler greyscale or RGB encodings often miss. Previous datasets such as MalNet \cite{Freitas}, Virus-MNIST \cite{Noever}, and MalImg \cite{Musaev} provide benchmarks but have limitations: MalNet’s byte-to-location colour mapping lacks obfuscation resilience, Virus-MNIST uses only the first 1,024 bytes, and MalImg’s small size risks overfitting. MalVis fills these gaps using a large AndroZoo sample \cite{Allix} and robust labelling via Euphony and VirusTotal, improving multiclass classification accuracy. Challenges remain, including class imbalance and visual similarity between malware families. Although undersampling and ensemble methods help, limitations in dataset diversity, interpretability, and scalability persist. Future work should enhance semantic feature extraction and adapt visualisations to capture malware behaviours beyond the current encoders.

\subsection{Feature-Rich Malware Datasets for Detection}
Borah et al., \cite{Borah} developed two detailed datasets: TUMALWD for Windows and TUANDROMD for Android. Their multi-phase framework includes data collection, analysis, and feature engineering. For Windows, honeynets capture binaries and network traffic, followed by sandbox dynamic analysis to extract API calls and network features \cite{Ghanem}. For Android, static analysis extracts permission- and API-based features from a large set of malware and benign apps. These recent datasets include thousands of samples and hundreds to thousands of features, addressing limitations of older datasets. However, both exhibit class imbalance, potentially biasing detection models. Reliance on sandboxing and static analysis may overlook sophisticated evasion and dynamic malware behaviours. Additionally, standardised public benchmarks are lacking for comparative evaluations. Despite these challenges, Borah et al., \cite{Borah} provide platform-specific, feature-rich resources, highlighting the need for balanced datasets, incorporation of real-time dynamic features, and universal standards in malware dataset creation. Their work emphasises the value of continuous dataset renewals against evolving threats \cite{Basnet}.

\subsection{Dataset for Malware Detection Research}
Sadek et al., \cite{Sadek} present over 4,600 memory snapshots from compromised Windows 10 VMs using obfuscation tools like Metasploit encoders, Shellter, Hyperion, and PEScrambler. They created encoded reverse shells to simulate advanced malware evasion, providing a valuable resource for machine learning detection. The dataset includes detailed labels such as process lists and memory maps, enabling forensic analysis. Sadek et al., \cite{Sadek} emphasise its support for cross-obfuscation testing and robustness evaluation against code obfuscation. However, scaling is difficult due to the cost of creating detailed snapshots, and the dataset may not reflect emerging malware types. Many malware datasets focus on Windows, limiting their generalisability. While Sadek et al., \cite{Sadek} partially address this, broader platform diversity and real-world complexity remain needed. Their research improves dataset quality and utility, but gaps persist in covering evolving threats and supporting cross-platform malware analysis.

\subsection{Realistic Malware Datasets from API Sequences}
Lu et al., \cite{Lu67} present a comprehensive malware variant dataset alongside API call sequences, addressing challenges in obtaining runnable, realistic obfuscated samples. They generated variants via binary rewriting obfuscation on PE files, including C/C++ and C\# malware, ensuring operational integrity. This contrasts with earlier methods that altered API sequences without guaranteeing executability, enhancing dataset authenticity. Using Cuckoo Sandbox, Lu et al., \cite{Lu67} dynamically extracted API call sequences under obfuscation, producing two datasets: one clear and one hidden, with over 9,000 and 8,000 sequences, respectively, increasing data variety. Their work supports robust detection models like BERT combined with TextCNN and adversarial training against obfuscation. Limitations include exclusive reliance on API sequences, ignoring parameters like timestamps, and challenges in slicing sequences that may lose critical behaviours or add noise. The adversarial generation method (FGM) is also suboptimal. Despite this, Lu et al., \cite{Lu67} fill gaps by combining executable variants with dynamic behaviour but suggest future work to enhance features and dataset construction.

\section{Research Methodology}
\label{sec:C}
\subsection{Experimental Environment Setup}
The experimental environment was designed using a virtual machine infrastructure to ensure controlled and isolated testing conditions. The infrastructure consisted of multiple virtual machines configured with different operating systems, allowing for a comprehensive analysis of malware behaviour across diverse platforms. An isolated network was established to prevent any unintended spread of malware beyond the testing environment, thereby maintaining safety and containment throughout the experiments. This isolation also allowed precise monitoring of network traffic and system interactions without external interference. Furthermore, the virtual machines were regularly reset to their clean states between infection trials to maintain data consistency. The use of virtualisation enabled rapid deployment and reconfiguration of the environment, thereby increasing the flexibility of the experimental setup. This well-designed environment created a safe and consistent base needed to take reliable memory snapshots and make sure the following analyses were accurate.

\begin{figure*}[h!]
\centering
\includegraphics[scale=0.25, angle=0]{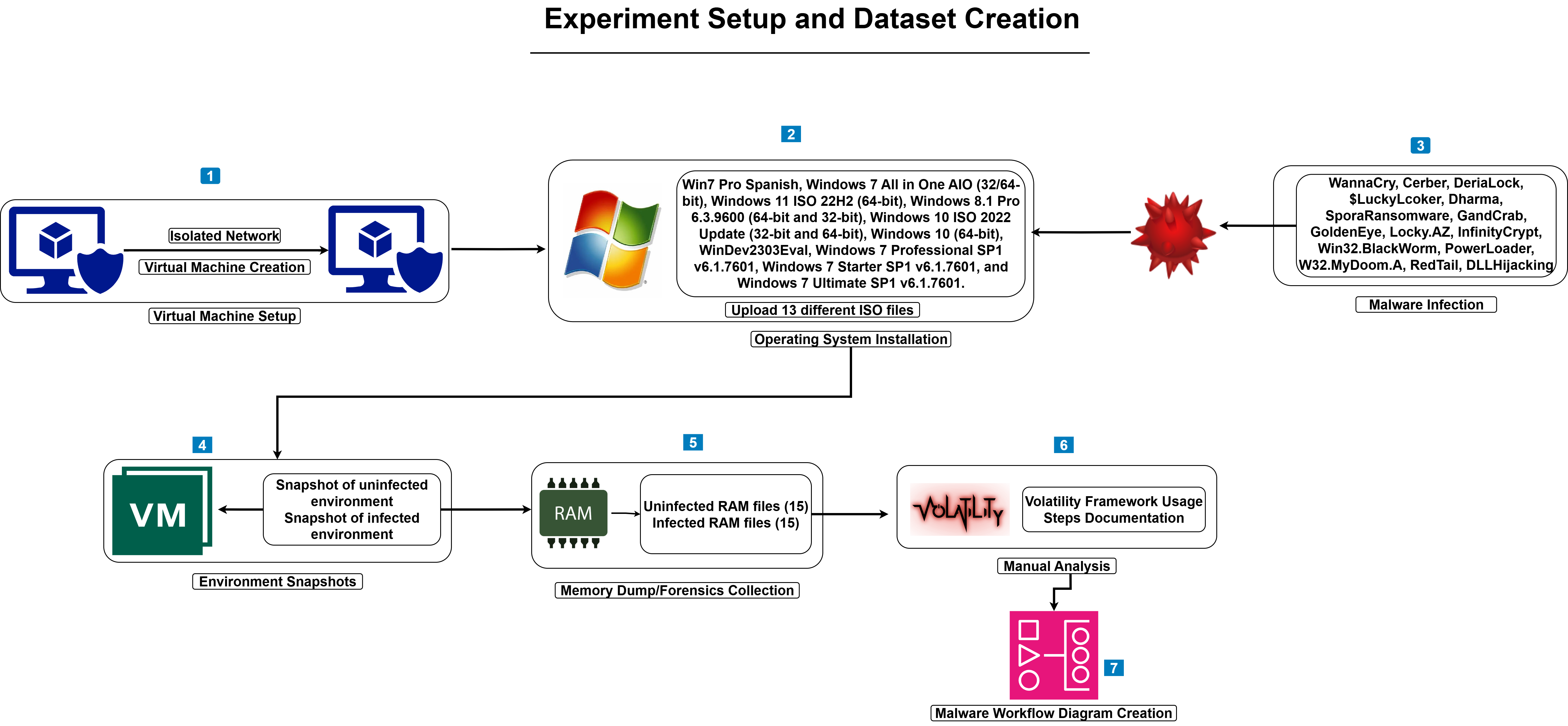}
\caption{Experimental Setup and Dataset Creation} \label{fig6}
\end{figure*}

\subsection{Selection of Malware and Operating Systems}
The selection of malware samples and operating system variants followed strict criteria to ensure relevance and diversity in the dataset. Malware was chosen based on its prevalence, diversity in behaviour, and potential impact on different OS platforms. The process involved sourcing well-documented malware families from reputable repositories, ensuring that the samples represented various attack vectors such as trojans, ransomware, and spyware. Concurrently, operating systems were selected to cover a broad spectrum of commonly used versions, including both legacy and modern releases, thereby reflecting realistic environments. This approach allowed for the systematic evaluation of malware effects across different system architectures. The careful pairing of malware and OS variants provided a robust foundation for generating meaningful data and capturing diverse infection scenarios, which was critical for comprehensive forensic and behavioural analyses.
\begin{table}[h!]
\centering
\footnotesize
\caption{\textit{Malware Names Against Windows Operating Systems, First Seen, and Category}}
\begin{tabular}{|l|l|l|l|}
\hline
\textbf{Malware Variant} & \textbf{Windows OS} & \textbf{First Seen} & \textbf{Category} \\
\hline
1 - PowerLoader & Windows 7 & 2010 & Trojan \\ 
2 - BlackWorm & Windows 7 & 2005 & Worm \\ 
3 - WannaCry & Windows 7 & 2017 & Ransomware \\ 
4 - W32.MyDoom.A & Windows 7 & 2004 & Worm \\ 
5 - Cerber & Windows 7 & 2016 & Ransomware \\ 
6 - Dharma & Windows 8.1 & 2016 & Ransomware \\ 
7 - LuckyLcoker & Windows 8.1 & 2016 & Ransomware \\ 
8 - SporaRansomware & Windows 10 & 2017 & Ransomware \\ 
9 - GandCrab & Windows 10 & 2018 & Ransomware \\ 
10 - GoldenEye & Windows 10 & 2016 & Ransomware \\ 
11 - InfinityCrypt & Windows 10 & 2020 & Ransomware \\ 
12 - Locky.AZ & Windows 11 & 2016 & Ransomware \\ 
13 - DeriaLock & Windows 11 & 2020 & Ransomware \\ 
14 - DLLHijacking & Windows 11 & 2023 & Injection Malware \\ 
15 - RedTail & Windows 11 & 2024 & Crypto Malware  \\ 
\hline
\end{tabular}
\end{table}

\subsection{Infection Process and Data Collection}
The infection process was designed to capture both clean and infected memory snapshots to enable effective comparison and analysis. Initially, clean RAM snapshots were taken from each virtual machine in its uninfected state to serve as a baseline. Subsequently, malware samples were introduced following controlled infection procedures tailored to each malware type. After allowing sufficient time for the malware to execute and manifest its behaviour, infected RAM snapshots were captured. The data collection process ensured consistent timing between infection and snapshot acquisition to standardise the dataset. The collected snapshots included various types, such as full memory dumps and selective memory region captures, with file sizes varying according to system configuration and the extent of infection. In total, the dataset comprises 30 clean and infected live memory dumps, carefully organised to facilitate analysis. This procedure ensured high-quality data that accurately reflected the memory state changes caused by different malware infections.

\subsection{Ethical and Legal Considerations}
Ethical and legal considerations were central to the research design to ensure full compliance with relevant standards and safeguard privacy. The data collection processes adhered to INTERPOL’s Data Protection Framework, emphasising the responsible handling of sensitive information without altering the integrity of the memory snapshots. Maintaining the original state of the snapshots was essential for preserving their forensic validity. We ensured legal compliance by adhering to applicable regulations on malware use, data handling, consent, and intellectual property. The experimental environment was strictly controlled to prevent any accidental spread of malicious code. Ethical approval for the research was awarded by the London Metropolitan University Ethics Board, confirming that all procedures met institutional and legal requirements. These measures ensured the research upheld high ethical standards while producing reliable forensic data. As a result of balancing the need to protect data with the importance of keeping evidence unchanged, the method used ensured that the research was both safe and ethical, following international guidelines.

\section{Significance Of Dataset Quality For Machine Learning Models}
\label{sec:D}
The forensic analysis used established tools to ensure accuracy and reliability of memory snapshots. Specifically, the Volatility Framework extracted forensic artifacts from clean and infected snapshots, automating detection of processes, network connections, injected code, and key malware indicators. Manual inspection confirmed artifacts and data integrity. This combined approach enabled cross-validation, improving dataset quality. Observations revealed clear differences between clean and infected states, including anomalous processes and network patterns. The dataset showed consistent snapshot sizes, diverse malware behaviours, and OS-specific artefacts, validating its use for malware detection. Additionally, machine learning success depends on high-quality, varied, and correctly labelled datasets. Collecting data from different malware types and operating systems improves AI-based detection and response.

\section{Documentation, Replicability, And Dataset Integrity}
Comprehensive documentation was essential to support the replicability and integrity of the dataset. We systematically maintained detailed records of the experimental setup, which included virtual machine configurations, malware samples, infection timelines, and snapshot acquisition procedures. This thorough documentation enables other researchers to reproduce the experiments and verify findings independently. Furthermore, version control and standardised naming conventions were implemented to accurately track the data's provenance and modifications. Data integrity was preserved through the use of cryptographic checksums and secure storage practices, ensuring that snapshots remained unaltered from capture to analysis. The importance of such measures lies in fostering transparency and trust in the dataset, which is critical for its adoption within the research community. As a result of combining documentation with strict integrity checks, the study provides a reliable resource that supports consistent, repeatable investigations into malware behaviours and digital forensic techniques.

\section{Challenges and Limitations}
During dataset formulation, several challenges required careful mitigation. A key issue was containing malware within the virtual environment to prevent unintended spread or damage, addressed by strict network isolation and frequent environment resets. The diversity of malware behaviours makes standardising infection procedures and snapshot timing complicated, necessitating adaptive protocols to capture relevant memory states. Additionally, virtualisation’s inherent constraints may not fully replicate hardware-specific behaviours on physical machines, potentially affecting the generalisability of results. The dataset’s focus on selected malware families and operating systems also limits coverage, possibly omitting emerging threats or less common OS variants. Despite these challenges, rigorous controls and validation steps minimised their impact. Acknowledging these limitations is vital for accurate interpretation and guiding future expansions to improve dataset coverage and realism.

\section{Expanding Dataset Utility for Generalised AI and Education}
This dataset not only advances malware detection and agentic AI, but it also broadens its use for generalised artificial intelligence and education. By including multiple operating systems, benign activities, and simulated user workloads, it captures a wide range of real-world system behaviours. Additionally, time-series memory snapshots synchronised with multimodal data such as network traffic, system call traces, and logs help AI models understand dynamic system states and transitions. As a result, the dataset helps create environments for reinforcement learning where the states show detailed pictures of the system, actions are linked to security or regular tasks, and rewards indicate how well the system detects issues and stays stable. Multi-level annotations improve interpretability, aiding AI training and student learning. Moreover, a modular dataset generation toolkit enables customised malware types, operating systems, and data modalities to meet diverse research and teaching needs. Baseline AI models and teaching materials accompany the dataset to encourage practical learning and wider adoption.

\section{Future Directions For Agentic And Hybrid Ai In Malware Detection}
The dataset represents a major advance in digital forensics and malware research by offering a comprehensive, well-validated resource for memory-based analysis across multiple malware families and operating systems, enabling a detailed study of diverse infection scenarios and behaviours. The combination of automated and manual validation enhances its reliability and forensic value while promoting the reproducibility and collaboration essential for cybersecurity progress. Still, there is scope for improvement by expanding to include more malware types, newer OS versions, and physical hardware environments to better mirror real-world conditions. Future enhancements could integrate dynamic behavioural logs and network traffic data to provide a more holistic view of malware activity that would aid countermeasure development. Building on this foundation, research may develop AI agents using multimodal data, combining symbolic reasoning with deep learning to improve explainability and trustworthiness. Crucially, continual learning and online adaptation will empower these agents to effectively counter zero-day exploits and evolving threats, advancing resilient, autonomous cybersecurity defences.

\section{Conclusion}
\label{sec:I}
This study presents a robust and carefully customised dataset designed to support memory-based malware analysis and incident response research. Key contributions include creating a secure testing environment, selecting diverse malware types and operating systems, and acquiring detailed memory snapshots from both clean and infected systems. Rigorous validation through automated tools and manual inspection ensures high data quality and reliability. Additionally, comprehensive documentation and integrity measures promote replicability and trustworthiness, encouraging wider adoption in the research community. The dataset is organised in such a way that it is easy to use with advanced machine learning and AI systems, creating great chances to improve how we detect and respond to malware. Ethical and legal considerations are thoroughly addressed, ensuring privacy and compliance throughout the research process. Finally, this dataset provides a reliable foundation for future cybersecurity research and innovation by capturing intricate malware behaviours at the memory level. Its availability will enhance cyber incident response and support the advancement of automated malware detection.

\section*{Data Availability}
The full data set is available at IEEE Dataport at \hyperlink{https://dx.doi.org/10.21227/kg5b-nf37}{https://dx.doi.org/10.21227/kg5b-nf37}

\end{document}